# COLLECTIVE INTELLIGENCE QUANTIFIED
# FOR COMPUTER-MEDIATED GROUP PROBLEM SOLVING

Dan Steinbock, Craig Kaplan, Marko Rodriguez, Juana Diaz, Newton Der, Suzanne Garcia

## ABSTRACT

**Collective Intelligence (CI) is the ability of a group to exhibit greater intelligence than its individual members. Expressed by the common saying that "two minds are better than one," CI has been a topic of interest for social psychology and the information sciences. Computer mediation adds a new element in the form of distributed networks and group support systems. These facilitate highly organized group activities that were all but impossible before computer mediation. This paper presents experimental findings on group problem solving where a distributed software system automatically integrates input from many humans. In order to quantify Collective Intelligence, we compare the performance of groups to individuals when solving a mathematically formalized problem. This study shows that groups can outperform individuals on difficult but not easy problems, though groups are slower to produce solutions. The subjects are 57 university students. The task is the 8-Puzzle sliding tile game.**

**Keywords:** collective intelligence, group problem solving, computer supported cooperative work, group decision support systems

## Introduction & Background

Collective Intelligence (CI) can be defined as the ability of a group to produce better solutions to a problem than group members could produce working individually [1].

Many human institutions are based on the belief that "two minds are better than one." It's the reason why democracies hold popular elections, why organizations use committees, and why business relies so much on meetings. Historically, Collective Intelligence is the motivation behind all forms of group problem solving—since the birth of collaboration.

The Social Psychology literature offers a number of empirical studies demonstrating the phenomenon of CI (e.g. [2], [3], [4], [5], [6]). Results in these studies generally indicate that group solutions are at least as good as the average quality of individuals' solutions.

For example, Yetton and Bottger [5] find group solutions to be superior in quality to average individual solutions for multi-part judgment problems and equal to the quality of the group's best member. Johnson and Johnson [6] find group solutions to be equal to or better than the best member's solutions when individual decision making precedes group discussion.

With the advent of information technologies such as local area networks, real-time digital communications and distributed software, researchers began exploring group problem solving in computerized network environments. For example, considerable research has been done on Computer Supported Cooperative Work (CSCW) and its sub-field Group Decision Support Systems (GDSS).

GDSSs combine communication, computer and decision technologies to support problem formulation and solution for groups [7]. In outlining their foundation for GDSS research, DeSanctis and Gallupe identify task type, group size and the presence or absence of face-to-face interaction as key variables [7]. Much research has studied the effect of these factors on the performance of computer-mediated groups, many comparing groups supported by computer to ones using traditional face-to-face methods (e.g. [8], [9], [10], [11]).

However there has been a lack of research comparing group to individual problem solving when computer-based decision support systems are provided to both [12]. In other words, though the CI phenomenon has been established in the Social Psychology literature, more work is needed to extend understanding into the digital realm of Computer Science.

Though there have been attempts towards filling this gap, the research to date has focused primarily on business-relevant qualitative tasks such as brainstorming [9], preference reconciliation [8] and multi-part judgment problems [12].

To date, little research has been published addressing whether CI occurs in computer-supported groups solving highly-structured, formalized problems (e.g. computer chess, Traveling Salesman). In contrast, well-defined problem solving tasks have been a primary area of research for Artificial Intelligence work on multi-agent algorithms [13].

Since network environments of the future will likely consist of both human and AI agents, it is important to explore whether groups of humans exhibit the phenomenon of CI when working on the types of highly-structured tasks where AI problem solving agents excel.

A better understanding of human performance on these types of problem solving tasks is a first step towards the study of problem solving by groups composed of both human and non-human agents.

This paper presents findings from an experimental study of computer-mediated group problem solving applied to a formalized task.

First we define the specific goals of our study, how we chose to achieve them, and the research questions we set out to answer.

Next, we enumerate our main hypotheses and describe the experiment and software system we designed to test them.

This is followed by a report of results we obtained during the experiment. We establish that the phenomenon of CI applies to groups of humans solving highly structured tasks in a network environment.

Next we discuss some of the factors that affect the degree of CI exhibited.

Finally, we consider the implications of this research for future directions in the study of Collective Intelligence.

### Research Focus

The goal of this research project is to investigate CI under well-defined laboratory conditions. Specifically, we desire quantitative measures of performance. To achieve this we take a novel approach to the study of computer-mediated group problem solving.

By encapsulating the process within a formalized software framework, all decision-making behavior can be recorded by the system. This allows a more rigorous, analytical model of human problem solving than previously possible.

To establish the CI phenomenon within this framework, we decided that a simple, mathematically formalized problem would be most suitable. The main benefit of this approach is that objective performance metrics (e.g. of solution quality, thinking time) become readily available that are absent from the study of other task types.

For example, solutions to a formalized maze problem can be objectively rated by path length. On the other hand, choosing which supplies would be "most useful" for surviving on the moon cannot be objectively measured—only compared to the opinions of experts [5].

For our study we chose the well-known 8-Puzzle sliding tile game. It fulfilled our need for a simple, formalized problem for which the solution space is pre-defined and for which mathematically objective metrics exist for solution quality and problem difficulty.

Implemented within our software framework, this approach provides the following benefits:

- Decision-making is isolated apart from solution generation
- Problem difficulty is objectively quantifiable
- Solution quality is objectively quantifiable
- Decision time is objectively quantifiable.

Solving this kind of formal problem is classified as an "intellective task" according to McGrath's model of group task types [14]. It involves finding a correct answer for which there exists an objective criterion of correctness [7].

For computer-mediated groups, intellective decision tasks can be efficiently solved with the exchange of specific facts across a low-bandwidth communications medium [15].

For groups attempting to collectively solve the 8-Puzzle, each person need only make a series of choices (votes) from a small set of alternatives. A system supportive of this task would leave this as the only cognitive burden on its users. All other needs, such as problem representation, vote-aggregation, and decision execution, would be met by the computer system.

We consider this an "algorithmic" approach to group problem solving.

Such a system would fall into the Level 3 class of Group Decision Support Systems as defined by DeSanctis and Gallupe [7]. They define a tri-leveled taxonomy of GDSSs based on the degree of computer involvement in decision-making processes. The higher the level, the more dramatic the intervention into the group's natural (unsupported) decision process.

Level 3 systems impose deliberate communication patterns onto the human decision-making process. For example a Level 3 system might enforce the rules of parliamentary procedure, or make possible the kind of highly-structured decision-making our framework calls for.

To this end we designed and implemented a distributed software system based around the 8-Puzzle. This system, which we named *Sliders*, was created to facilitate experimental data-gathering and answer the following research questions:

- How does solution quality of computer-mediated groups compare to individuals supported by comparable software?
- How does group performance compare to the average individual?
- How does group performance compare to the best individual?
- How does solution time vary according to similar comparisons?
- If differences in solution quality and/or time are exhibited, which experimental variables correlate to the difference?
- What effect does problem difficulty have on group and/or individual solution quality?

### Experiment Design

The following hypotheses were constructed based on the findings of past investigations into CI (see Introduction) and our own intuitions regarding the research questions stated above.

**Hypothesis 1a:** Group solution quality is higher than average individual solution quality at all difficulty levels.

**Hypothesis 1b:** Group solution superiority over the average individual is greater on hard problems than on easy problems.

**Hypothesis 2:** Group solution quality is better than the best individual's solution quality at all difficulty levels.

**Hypothesis 3:** Group solution time is greater than average individual solution time at all difficulty levels.

To compare the performance of collaborative groups with the performance of individual subjects, we used a within subjects experimental design with two main experimental conditions. 57 undergraduate and graduate students were assigned to one of two groups.

In the first experimental group (the Group Condition), subjects solved a series of 8-Puzzle problems working together in a collaborative group. Under a 30-second time constraint, each subject voted for his or her favorite next move at each step of problem solving. The software then executed the move which received the most votes. Subjects played as many games as they could successfully solve within a 30 minute time period. Further, the sequence of problems was arranged so that each problem was slightly more difficult than the previous problem. Thus, subjects solved problems of increasing difficulty as the experiment progressed.

In the second experimental group (the Solo Condition), subjects solved a series of 8-puzzle problems that increased in difficulty -- just as the Group Condition did. However this time, the subjects worked alone. That is, each member of the solo group chose his or her favorite next move at each stage of problem solving and the software immediately made this next move. Thus while the Group Condition performance reflected the efforts of many individuals voting at each step, the Solo Condition performance reflected individual problem solving efforts.

The within subjects design meant that the same players served in both experimental groups. Half of the players solved problems first in the Solo Condition and then later in the Group Condition. Other players solved problems first in the Group Condition and then later in the Solo Condition. This counterbalanced design was meant to control for possible learning as subjects gained more experience solving the problem.

### System Implementation

*Sliders* is a particular instantiation of the well-known 8-Puzzle tile sliding game. Along with its cousins the 15-puzzle and 24-puzzle, this problem is based on simple rules yet proves very difficult in practice. Because of its combinatorially large problem space, it has for many years been used as a testbed for heuristic search techniques [16]. The N*N extension of the 8-Puzzle is known to be NP-hard [17].

The 8-Puzzle consists of a 3x3 grid of eight numbered tiles from 1 to 8, and one empty slot. Given a scrambled initial layout of tiles, the player must rearrange the tiles into a goal configuration by sliding tiles orthogonally into the current empty slot.

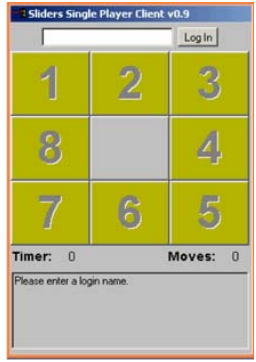

*Figure 1: The Sliders client interface in its goal state.*

For the goal state we adopt Korf's [18] convention: tiles are in numerically increasing order clockwise around the grid, with the empty slot in the center (see fig. 1). Degree of difficulty for a given initial state is defined as the minimum (optimal) number of moves required to reach the goal state. For instance, fig. 2 shows a difficulty 2 game state and fig. 3 is of difficulty 8.

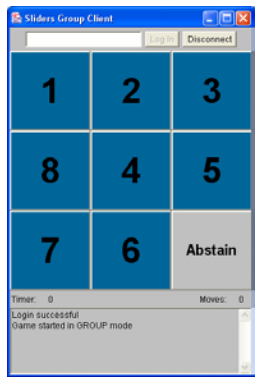

*Figure 2: A Sliders board 2 moves from the goal.*

*Sliders* is a client/server network application. A centralized server contains a database that logs all the behaviors of all participants. There are two operating modes: a single-player mode (for the Solo Group condition) and a group-player mode (for the Group Condition).

In single-player mode, a player logs into the central server. The server then sends a new unsolved puzzle layout, and the user begins playing by selecting tiles adjacent to the empty tile. The player has 30 seconds to make a move, as shown by a countdown timer. When a move is made, the board and move counter are updated, and the timer is reset. If a move is not made within the allotted time, the move counter is still incremented and the timer is reset. After the user solves the puzzle, there is a five second delay before the server gives a new puzzle of greater difficulty. The player attempts to solve as many puzzles of increasing difficulty over a 30 minute period.

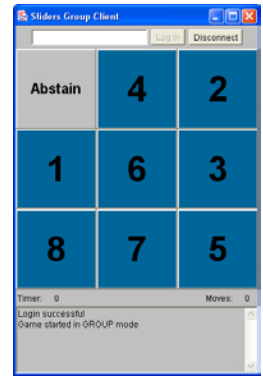

*Figure 3: A Sliders board 8 moves from the goal.*

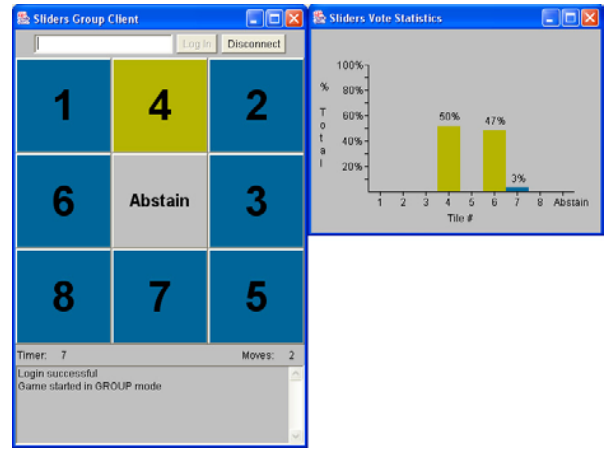

*Figure 4: The Sliders Group client interface with feedback display showing vote statistics.*

Group mode is similar to single-player mode except that multiple players collaborate by voting on the next move to make. Players are presented with a chart that displays the amount of votes each tile has received during that move round (see fig. 4). This feedback mechanism allows group players to see the voting behavior of their peers. Players are able to change their vote any number of times during the 30 second move round. When all votes have been cast or 30 seconds have elapsed, whichever comes first, the current voting round ends. The tile with the most votes is the tile that is moved and the next vote round proceeds with the new game state. When the puzzle is returned to its goal configuration, the server then repeats in the same manner, waiting five seconds before giving a new puzzle. The group attempts to solve as many puzzles as possible within the given 30 minute period.

## Experimental Results

This section presents the data obtained from our experiment, organized around the hypotheses defined above.

The dependent variables are solution quality (of the Group Condition, average Solo Condition member and best Solo Condition member) and time to solution (of the Group Condition and average Solo Condition).

Solution quality for a given initial puzzle state is measured against the known optimal distance to the goal state. Data has been normalized, i.e. Solution Quality = Actual Number of Moves – Known Optimal Distance. Thus an optimal solution would score zero.

Time to solution is calculated as the number of seconds elapsed from the initial appearance of a puzzle to the execution of its solving move.

Average and Best players are calculated by summing up the amount of moves it takes every solo player to complete puzzles at each difficulty level. Dividing by the number of players yields the average. Choosing the minimum identifies the best player.

In the tables that follow, problem difficulty is divided into two categories: Easy (optimal distance 1-8) and Hard (9-16).

Hypothesis 1a states that group solution quality is higher than average individual solution quality at all difficulty levels. Though the Group condition solves problems in fewer moves overall than the Solo condition, this difference is not statistically significant (see table 1). Thus hypothesis 1a is rejected.

| Difficulty | Avg. Solo | Group |
|---|---|---|
| Easy | 11.016 | 5.279 |
| Hard | 41.137 | 10.5 |
| Overall | 24.615 | 7.182 |

*Table 1: Mean # of moves, Group x Avg. Individual x Problem Difficulty.*

Hypothesis 1b states that group solution quality should exceed average individual solution quality to a greater degree on hard problems than easy problems. When we examine only the most difficult problems, we find that the group condition is better than the individual condition (see table 1) to a statistically significant degree (t = 3.1, p < .05). Thus hypothesis 1b is accepted.

| Difficulty | Best Solo | Group |
|---|---|---|
| Easy | 4 | 4.7 |
| Hard | 13.833 | 10.078 |
| Overall | 8.538 | 7.182 |

*Table 2: Mean # of moves, Group x Best Individual x Problem Difficulty*

Hypothesis 2 states that group solution quality is better than the best individual's solution quality at all difficulty levels. This hypothesis is rejected because the performance difference over all levels is not statistically significant (see table 2). The same is true for easy and hard difficulty levels taken separately.

| Difficulty | Avg. Solo | Group |
|---|---|---|
| Easy | 24.374 | 142.189 |
| Hard | 56.653 | 202.038 |
| Overall | 52.307 | 180.089 |

*Table 3: Mean solution time in seconds, Group x Avg. Individual x Problem Difficulty.*

Hypothesis 3 states that group solution time is greater than average individual solution time at all difficulty levels. We find that the group condition takes more time at every difficulty level than the individual condition (see table 3). This finding is statistically significant (t = 6.46, p < .05) so the hypothesis is accepted.

## Discussion

We defined Collective Intelligence as the ability of a group to produce better solutions to a problem than group members could produce working individually. The intent of our experiment was to establish the phenomenon of CI in a formalized problem domain.

Our most significant finding is that CI *can* be exhibited by groups of humans solving highly structured intellective tasks in a network environment. Specifically, our experimental results show that group solution quality is significantly greater than the average individual solution quality when solving hard problems but not when solving easier problems.

While in general the group outperforms the average individual on the easier difficulty levels, the difference is statistically significant only for the harder problems.

Interestingly, our findings also indicate little difference between group solution quality and the best individual's solution quality—for all difficulty levels. This result is especially noteworthy with respect to Gallupe's [12] finding that GDSS-supported groups are significantly outperformed by their best individuals. Since Gallupe's study used a multi-part judgment task, the possibility is raised that task type is a factor determining the relative performance of a group and its best individual.

As expected, group solution time is significantly greater than the average individual solution time for all difficulty levels.

Several possible explanations could account for these findings:

One social psychological explanation for CI comes from the interactional theory of group decision-making. This posits that groups perform better than individuals because greater resources are available to each group member, thus motivating performance, creativity and error-correction [2]. In our computer-based experiment, group interaction occurs via the voting statistics feedback display. Cognitive feedback in GDSSs has been shown to be an effective mechanism for facilitating convergence among group members [19]. We plan to compare feedback versus no-feedback in a future experiment.

Another explanation raised in our post-experiment analysis is that constraints imposed by our system may influence group solution time. Because the time spent at each decision point is the time spent by the slowest individual (or the 30 second limit, whichever is less) there are times when faster group members are left waiting for slower members to make a decision. In contrast, individuals working alone never have to wait before moving to the next decision point. Thus the group could potentially move much quicker if different constraints are imposed by the system (e.g. executing a decision when a *majority* of members have made a decision instead of *all* members).

This factor may also influence group solution quality due to the forced introduction of additional thinking time into the problem solving process.

Our last explanation for CI derives from AI research into multi-agent problem solving. Agent-based algorithms modeled on the group behavior of social insects have been highly successful when applied to such problems as task allocation [20] and shortest path optimization [21]. These insects exemplify the CI property by exhibiting a sophisticated, emergent intelligence out of a large group of relatively unintelligent, autonomous individuals.

For example, ants foraging for food leave a pheromone trail which other ants can detect and, depending on how thick the deposit is, follow. Over time, the more popular paths are reinforced and gain a stronger pheromone scent while the unpopular paths slowly lose their scent. Thus the network of trails becomes a shared statistical indicator of the group's overall behavior—just like the voting graph feedback in the *Sliders* interface.

Ant Algorithms have been shown to be extremely effective on combinatorial optimization problems such as the Traveling Salesman Problem and Graph-Coloring Problem [21].

Perhaps future extensions of the present research will discover a new class of well-defined Human Algorithms that exemplify the CI property.

Moreover, since future network environments will likely incorporate both human and non-human intelligence, it is important to investigate human problem solving within the kind of highly structured domain where non-human agents currently excel.

Formalized problem solving systems like *Sliders* create a common ground for human and artificial intelligence. This branch of research is a first step towards ultimately investigating how groups composed of both human and non-human agents might collaborate together most effectively.

## Conclusion

This research contributes to the body of literature comparing group versus individual problem solving and attempts to elucidate the factors which combine to exhibit the Collective Intelligence phenomenon. It also furthers understanding of Group Decision Support Systems when applied to intellective tasks and highlights issues of computer-mediation that may affect group decision quality and time.

Given the limited scope and lack of antecedents to this research, more extensive studies and laboratory testing with different decision support systems are needed before computer-mediated group problem solving of highly structured tasks is fully understood. We hope future researchers will replicate our findings and further current understanding of Collective Intelligence.